\documentclass[showpacs,preprintnumbers,amsmath,amssymb,a4paper]{revtex4}
\usepackage{graphicx}% Include figure files
\usepackage{dcolumn}% Align table columns on decimal point
\usepackage{bm}% bold math
\usepackage{epsfig}
\usepackage{colordvi}

\newcounter{fig}

\newcounter{One}
\newcounter{Two}
\newcounter{Three}
\def\beq{\begin{equation}}
\def\eeq{\end{equation}}
\newcommand{\gesim}{\lower.7ex\hbox{$\;\stackrel{\textstyle>}{\sim}\;$}}
\newcommand{\lesim}{\lower.7ex\hbox{$\;\stackrel{\textstyle<}{\sim}\;$}}

\begin{document}

\title{Resonant flavor conversion of supernova neutrinos and neutrino parameters}

\author{Shao-Hsuan Chiu}
\altaffiliation[]{schiu@mail.cgu.edu.tw}
\affiliation{Physics Group, CGE, Chang Gung University \\
Kwei-Shan 333, Taiwan}

%\date{ }

\begin{abstract}

The unknown neutrino parameters may leave detectable signatures
in the supernova (SN) neutrino flux.
However, even the contribution from the MSW flavor transition
alone could cause
ambiguity in the interpretation to the neutrino signals
because of the uncertain local density profile of the SN matter
and the model-dependent SN neutrino spectral parameters.
A specific parametrization to the unknown local density profile is proposed
in this work,
and the contribution from the standard MSW effect is investigated
through a multi-detector analysis of the SN neutrinos.
In establishing the model-independent scheme, 
results based on the existing spectral models are included.
The limitation of the analysis is also discussed.

\end{abstract}


\maketitle

%%%%%%%%%%%%%%%%%%%%%%%% sec 1 %%%%%%%%%%%%%%%%%%%%%%%%%%%%%%%%%%%%%%%

\section{Introduction}

As a distinctive type of the neutrino source, 
the core-collapse supernova (SN) provides a
rich physical content that is lacking in the terrestrial environment.
With its unique production and detection processes, 
the neutrino burst from a SN has long been considered as
one of the promising tools for probing the unknown neutrino
intrinsic parameters~\cite{dighe:00,ref2,ref3,ref4,ref5,ref6,ref7}, %%%%%%%%%%%%% 
in particular, the neutrino mass hierarchy
and the tiny mixing angle $\theta_{13}$.
However, difficulties also arise from the complexity 
caused by the unavoidable
astrophysical uncertainties, which could lead to ambiguous interpretations
of the observed events. 

The existing paradigm for the neutrino MSW flavor conversion in a SN
has been established with the assumption of small neutrino self interaction.
However, a possible new paradigm has been shaped in recent years.
The argument is that, in addition to the standard MSW effect, the neutrinos
may encounter certain non-MSW effects in the SN environment when the neutrino
number density is extremely high.  These not so well-known effects may be independent of
the MSW effect and may introduce additional factors that alter the efficiency 
of the neutrino flavor transition.  The current consensus is that both types of 
effects could contribute to the flavor transition
and should all be included in the more complete analysis of the SN neutrino signals.

Even the standard treatment of the MSW flavor conversion 
alone is not immune from ambiguity.
The uncertain local density profile of the SN matter and the undetermined 
neutrino spectral parameters, such as the average energy and the luminosity
of each neutrino flavor, are the crucial factors involved in the MSW effect of 
the SN neutrinos.
However, an analysis based on the simple global density model: $\rho \sim r^{-3}$,
which is widely adopted in the literature, could lose the generality if
the variation of the 
local density shape near the resonance
is not taken into account.
With these uncertain factors, it is worth while to investigate 
how the contribution from the standard MSW effect should be modified
in analyzing the SN neutrino signals,
unless the overall contribution from the non-MSW effects is much greater
than that from the MSW effect.

%%%%%%%%%%%%%%%%%%%%%%%%%%%%%

The resonant neutrino flavor transition in a SN and in Earth would 
in principle give rise to observable signatures 
that reflect the properties of neutrino parameters.
The unknown local density profile of SN matter 
near the resonance takes part in the adiabaticity parameter of
the level crossing, and plays a role in the determination of 
transition probabilities.  
In fact, the adiabaticity of the level crossing could vary abruptly
with the local density profile in certain neutrino parameter space.
In addition, 
the knowledge to the primary spectrum
for each neutrino flavor is essential in assessing the efficiency
of neutrino flavor conversion.
Various mixing scenarios, which arise from the uncertain SN physics,
the possible neutrino mass hierarchies, and the uncertain magnitude
of $\theta_{13}$ lead to distinct neutrino survival probabilities.
An analysis of the observed neutrino signals should, in principle, be
able to single out the working scenario.

The promising features of the multi-detector experiments 
have been well recognized~\cite{ls:01,mirizzi,dighe:03,minakata:08,sf:08}. %%%%%%%
With the uncertain local density profile and the spectral parameters,
this work investigates the potential signatures 
that may be related to the contribution from the MSW effect in
a multi-detector analysis of the SN neutrinos. 
As suggested by the previous work~\cite{chiu06,chiu07},  %%%%%%%%%%%%%%%%
the consequences due to the uncertain local density profile near the resonance 
can be accounted for by the adoption of 
independent and variable power-law density functions
$\rho(r)=c_{k} r^{n_{k}}$.  However, the calculations in Ref. 13 and 14
are performed under the assumption of nearly constant $c_{k}$, 
which is similar to that adopted
in the usual analysis based on $\rho (r)= c r^{-3}$.
In the present work, the key improvement in dealing with the variable density
profile is the establishment of variable $c_{k}$, which reproduces
reasonable densities that agree with the numerical simulations
at a wide range of the radial location. 
With the more general parametrization of the density function, 
this work is further devoted to analyzing
the expected neutrino events at two  
water Cherenkov detectors. Focus is aimed at the
modified contribution from the MSW effect. 
Certain physical observables derived
from the expected event rates at the two detectors are proposed
as the discriminators of various transition scenarios for the MSW resonance. 
In searching for the model-independent
properties of the observables,
three existing SN neutrino spectral models proposed by
the Garching group~\cite{Ga,Gb} and the Lawrence Livermore group\cite{LL}
are adopted in the calculation.

%%%%%%%%%%%%%%%%%%%%%%%%% sec 2  %%%%%%%%%%%%%%%%%%%%%%%%%%%%%%%%%%%%%%%%%%%%%

\section{Uncertain density profile of SN matter}

There is no definite way of modeling
the practically unknown local variation of density profile.
Considering the narrow thickness of the resonant layer, as
compared to that of the whole scope of the SN matter distribution, 
the density profile with a form of variable power-law in each 
resonant layer would be a reasonable simplification that accounts
for the possible dynamical consequences. %%%%%%%% 
The essence of this approximation
is that the power characterizing the density profile in each 
resonant layer is allowed to vary independently: 
$\rho (r) \sim c_{l} r^{n_{l}}$ for the lower resonance layer, and
$\rho (r) \sim c_{h} r^{n_{h}}$ for the higher resonance layer,
where the factor $c_{k}$ denotes the magnitude of the profile $r^{n_{k}}$,
with $k=l$ or $h$.  

%%%%%%%%%%%%%%%%%%%%%%%%%%%%%%%%%%%%%%%
%%%%%%%%%%%%%%%%%%%%%%%%%%%%%%%%%%%%%%%%%%%%%%%

In this present work, the density profile in the resonance layer is parameterized as
\begin{equation}
\rho (r)=[\frac{c}{R_{0}^{3}}][\frac{r}{R_{0}}]^{n_{k}},
\end{equation}
where $c$ is a mass scale, and $R_{0}$ is a distance scale, which is
conveniently taken as the solar radius, $R_{0} \simeq 6.96 \times 10^{10}$ cm.
With the variable $n_{k}$, an important criterion for a reliable parametrization of the
density profile is that the predicted density
at a specific radial location agrees with that given by the numerical simulation, 
at least in order of magnitude. This suggests that in Eq.(1)
the variation of $n_{k}$ is accompanied by the variation of $c$ at 
a different location.  One may thus write the density profile in a more convenient form as
\begin{equation}
\rho (r)=[\frac{c_{0}}{R_{0}^{3}}][\frac{c}{c_{0}}][\frac{r}{R_{0}}]^{n_{k}},
\end{equation}
where $c_{0}$ is a constant mass scale, 
and $c_{0}/R_{0}^{3}$ is a reference density scale.
In the typical model with a fixed power, $\rho(r)=c r^{-3}$, 
the magnitude of $c$ varies weakly\cite{kuo88}: $10^{31}g<c<15 \times 10^{31}g$
in the density range $10^{-5} g/cm^{3} < \rho < 10^{12} g/cm^{3}$. 
For variable $n_{k}$ near a given location, however, the
mass scale $c$ may vary by several orders of magnitude so as 
to reproduce reasonable density at this specific radial location.  
Note that with this variable parametrization of the density profile,
the choice of $c_{0}$ does not alter the results since the variation
of the mass scale is represented by $c$. In this present analysis
we shall adopt the value $c_{0}=7.0 \times 10^{31}g$, which is simply
the mean value of $c$ in the typical density model.
The variables $c$ and $r$ may be rewritten as $c/c_{0}=X$ and $r/R_{0}=Z$, respectively,
and the density profile becomes
\begin{equation}
\rho (r)=[\frac{c_{0}}{R_{0}^{3}}] X Z^{n_{k}}.
\end{equation}
By fitting the predicted densities with the numerical results\cite{woosley-1,woosley-2,sato:03}
at different radial locations,  %%%%%%%%%%%%%
this parametrization leads to an approximate  
relation that regulates the variation of $c$, $r$, and $n_{k}$:
\begin{equation}
X \sim \frac{1}{2} Z^{-(n_{k}+3)}.
\end{equation}

%%%%%%%%%%%%%%%%%%%%%%%%%%%%%%%%%%%%%%%%%%%%%%%%%

As summarized in the following, the variation of
$n_{k}$ can lead to non-trivial effects on the crossing probabilities
at the resonance. 
One may first write the electron number density $N_{e}(r)$ for a typical 
core collapse SN as 
\begin{equation}
N_{e}(r)=\rho (r)[Y_{e}/m_{n}],
\end{equation} 
where the electron number per baryon $Y_{e}=1/2$ is adopted,
and $m_{n}$ is the baryon mass.  
The adiabaticity parameter for the resonant transition, 
defined as\cite{kuo89} 
\begin{equation}\label{eq:gamma1}
 \gamma_{k} \equiv \frac{\delta m^{2}_{ij} \sin^{2}2\theta_{ij}}
{2E \cos 2\theta_{ij}|\frac{1}{N_{e}} \frac{dN_e}{dr}|_{0}},
\end{equation}
becomes
\begin{equation}\label{eq:gamma2}
\gamma_{k} = \frac{1}{2|n_{k}|}[\frac{\delta m^2_{ij}}{E}]
   [\frac{\sin^2 2\theta_{ij}}{\cos 2\theta_{ij}}]
   [(\delta m^2_{ij}/E)\frac{\cos 2\theta_{ij}}
   {\sqrt{2} G_{F}(\frac{Y_{e}}{m_{n}})(\frac{c_{0}}{R_{0}^{3}})}]^{\frac{1}{n_{k}}} 
   R_{0} Z_{0}^{(1+\frac{3}{n_{k}})},
\end{equation}
where $\delta m^{2}_{ij} \equiv m^{2}_{i}-m^{2}_{j}$,
$\theta_{ij}$ is the mixing angle between the eigenstates 
$\nu_{i}$ and $\nu_{j}$, 
$G_{F}$ is the Fermi constant, $E$ is the neutrino energy, and
$Z_{0}=r_{0}/R_{0}$ indicates the location of resonance.
Note that $|(1/N_{e})(dN_{e}/dr)|_{0}$ is evaluated at the resonance,
where 
\begin{equation}
(N_{e})_{0}=\frac{\delta m^{2}_{ij} \cos 2\theta_{ij}}{2\sqrt{2}G_{F}E}.
\end{equation}
The adiabaticity parameter $\gamma_{k}$ appears in the level
crossing probability $P_{k}$ (with $k=l$ or $h$) as:
\begin{equation}\label{eq:pc} 
 P_{k} = \frac{\exp[-\frac{\pi}{2} \gamma_{k} F_{k}]-
 \exp[-\frac{\pi}{2} \gamma_{k} \frac{F_{k}}{\sin^{2} \theta_{ij}}]}
 {1-\exp[-\frac{\pi}{2} \gamma_{k} \frac{F_{k}}{\sin^{2} \theta_{ij}}]},
   \end{equation}
where $F_{k}$ is the correction factor to a non-linear profile.
The origin of Eq. (9) and the correction function $F_{k}$ can be found
in, $e.g.$, Ref. 22.
We shall adopt $\delta m^{2}_{21}=7.0 \times 10^{-5}$ eV$^{2}$, 
$|\delta m^{2}_{13}|=3.0 \times 10^{-3}$ eV$^{2}$, and $\sin^{2}2\theta_{12}=0.81$
in the calculation.  With a given radial location for the resonance,
it can be verified that
the adiabaticity parameter
$\gamma_{k}$, and thus the neutrino survival 
probability, vary weakly with the energy $E$,
as compared to the influences from the variations of
$n_{k}$ and $\theta_{13}$. 
This result suggests that
one may simply adopt the average neutrino energies, $e.g.$,
$E_{\nu_{e}}=12$ MeV and $E_{\bar{\nu}_{e}}=15$ MeV, in the following calculation
for the neutrino transition probability in SN.
The energy dependence would only appear in calculating the Earth effect.

%%%%%%%%%%%%%%%%% sec3 %%%%%%%%%%%%%%%%%%%%%%%%

\section{Survival probabilities and scenarios for neutrino parameters}

When the neutrinos arrive at Earth, 
the survival probabilities for $\nu_{e}$ and $\bar{\nu}_{e}$ 
are given respectively by 

\begin{equation}\label{eq:pe}
       P_{nor} = U_{e1}^{2} P_{l} P_{h} + U_{e2}^{2} (1-P_{l}) P_{h} +
                 U_{e3}^{2} (1-P_{h}),
\end{equation} 
   
\begin{equation}\label{eq:pa}
   \bar{P}_{nor} = U_{e1}^{2} (1-\bar{P}_{l}) + U_{e2}^{2} \bar{P}_{l},
\end{equation} 
for the normal hierarchy, and

\begin{equation}\label{eq:pei}
    P_{inv} = U_{e2}^{2} (1-P_{l}) + U_{e1}^{2} P_{l},
\end{equation}
   
\begin{equation}\label{eq:pai}
    \bar{P}_{inv} = U_{e2}^{2} \bar{P}_{l} \bar{P}_{h}+
                  U_{e1}^{2} (1-\bar{P}_{l}) \bar{P}_{h} + 
                 U_{e3}^{2} (1-\bar{P}_{h}),
\end{equation}
for the inverted hierarchy,
where $P_{h} (\bar{P}_{h})$ and $P_{l} (\bar{P}_{l})$ represent the
higher and the lower level crossing probabilities for
$\nu_{e} (\bar{\nu}_{e})$, respectively, and $U_{ei}$ is the element of the mixing matrix.

%%%%%%%%%%%%%%%%% fig1%%%%%%%%%%%%%%%%%%%

\begin{figure}
\caption{The rapid variation of $P_{h}$ as functions of $\sin^{2}2\theta_{13}$
for $n_{h}=-8$ and $n_{h}=-2$. \label{f1}} 
\centerline{\epsfig{file=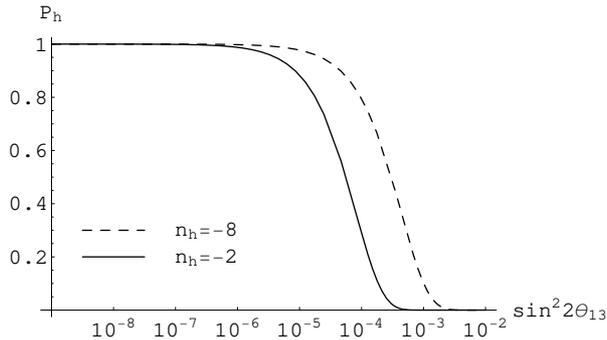,width=8 cm}}
\vspace*{8pt}
\end{figure} 

%%%%%%%%%%%%%%%%%%%%%%%%%%%%%%%%%%%%%%%%%%%%%%%%%%%%%%

In calculating the above level crossing probabilities,
it should be emphasized
that both the higher and the lower
level crossings occur in the $\nu$ sector for the normal hierarchy,
while the higher crossing occurs in the $\bar{\nu}$ sector 
and the lower crossing occurs in the $\nu$ sector if the mass hierarchy is inverted. 
One thus needs to calculate 
$P_{h}$ and $P_{l}$ in Eq.(10) for the normal hierarchy, and $\bar{P}_{h}$ in Eq. (13)
and $P_{l}$ in Eq.(12) for the inverted hierarchy at the individual resonance.
Note that since the antineutrinos do not
encounter the resonance at the location where the neutrinos cross the lower resonance,
the crossing probability $\bar{P}_{l}$ for both hierarchies in
Eq.(11) and Eq.(13) is vanishing: $\bar{P}_{l}=0$.
In addition, it can be verified that
the adiabaticity parameter $\gamma_{l}$ at the lower resonance
is very large, $\gamma_{l} > 10^{2}$.
This suggests that the lower resonance is adiabatic
for both the mass hierarchies: $\gamma_{l} \gg 1$, 
which leads to $P_{l} \sim 0$.

On the other hand, to show that the variation of $n_{h}$ and $\theta_{13}$
leads to non-trivial structures of $P_{h}$ in the $n_{h}-\theta_{13}$ parameter space,
we first estimate the typical density near the location of the higher resonance.
Using Eqs.(5) and (8), the density at resonance is given by
\begin{equation}
\rho_{res}=\frac{\delta m_{31}^{2} \cos 2\theta_{13}}{[2\sqrt{2}G_{F}E][Y_{e}/m_{n}]},
\end{equation}
which leads to $\rho_{res} \simeq 3 \times 10^{3} g/cm^{3}$ with small $\theta_{13}$.
This density corresponds to an approximate location of resonance 
$r_{0}/R_{0}=Z_{0} \sim 0.03$ 
according to the numerical results~\cite{woosley-1,woosley-2,sato:03}. 
It follows from Eqs.(7) and (9) that the crossing probability $P_{h}$  %%%%%%%%% eq#
is in general non-adiabatic ($P_{h} \sim 1$) for $\sin^{2}2\theta_{13}<10^{-5}$.
This leads to $P_{nor}=P_{inv} = |U_{e2}|^{2} \sim 0.3$, as can 
be checked using Eqs. (10) and (12).  In addition, 
it can be verified that $P_{inv}$ remains constant through out the
parameter space with $P_{inv} \sim 0.3$.  However, the probability $P_{nor}$ varies
with both $n_{h}$ and $\theta_{13}$ near $\sin^{2}2\theta_{13} \sim 10^{-5}-10^{-4}$  
since $P_{h}$ drops rapidly from $P_{h} \sim 1$ (non-adiabatic) to $P_{h} \sim 0$ (adiabatic)
within a narrow region of the parameter space.
It follows that $P_{nor}$ drops from $P_{nor} \sim 0.3$ to $P_{nor}= |U_{e3}|^{2} \ll1$.
This property distinguishes $P_{nor}$ from $P_{inv}$.
As an illustration, we show $P_{h}$ as a function of $\sin^{2} 2\theta_{13}$
for $n_{h}=-2$ and $n_{h}=-8$ in Fig. 1.

%%%%table 1%%%%%%%%%%%

 \begin{table}
\caption{The survival probabilities for $\nu_{e}$ and $\bar{\nu}_{e}$, 
the possible neutrino mass hierarchy, and the function $g(n_{h},\theta_{13})$
as predicted by the specific scenarios.
   The uncertainty in the power of the local density profile 
   can alter the predicted bound of
 $\theta_{13}$ through the function $g(n_{h},\theta_{13})$.}
 {\begin{tabular}{lllllllll}  \hline \hline

       &  & $P$ &   & $\bar{P}$  &    &  mass hierarchy   &  & $g(n_{h},\theta_{13})$    \\ \hline
$(a)$ &  & $|U_{e3}|^{2}=\sin^{2}\theta_{13}$ &   & $|U_{e1}|^{2} \simeq \cos^{2}\theta_{12}$  & & normal   &   & $>1$       \\
 $(b)$ &  & $|U_{e2}|^{2} \simeq \sin^{2}\theta_{12}$ &   & $|U_{e3}|^{2}=\sin^{2}\theta_{13}$  && inverted  & & $>1$   \\
 $(c)$ & & $|U_{e2}|^{2} \simeq \sin^{2}\theta_{12}$ &   & $|U_{e1}|^{2} \simeq \cos^{2}\theta_{12}$  & & both & & $<1$     \\
                        \hline
     \end{tabular}\label{ta1}}
 \end{table}

 %%%%%%%%%%%%%%%%%%%%%%%%%%%%%%%%%%%%%%%%
 
It is then reasonable to represent the narrow borderline between the adiabatic crossing
($P_{h} \sim 0$) and
the non-adiabatic crossing ($P_{h} \sim 1$) by the condition
$P_{h} = 1/2$, which implies (with small $\theta_{13}$)
\begin{equation}
\exp[-\frac{\pi}{2} \gamma_{h} F_{h}] \sim \frac{1}{2}.
\end{equation}
Equations (7) and (15) lead to $\frac{\pi}{2 \ln2}\gamma_{h} \equiv g(n_{h},\theta_{13}) \sim 1$, 
with
\begin{eqnarray}
  g(n_{h},\theta_{13}) & = &
  \frac{\pi}{4(\ln2)}
  \frac{1}{|n_{h}|}[\frac{|\delta m^2_{31}|}{E}]
   [\frac{\sin^2 2\theta_{13}}{\cos 2\theta_{13}}]  \nonumber  \\
   & & \times [(|\delta m^2_{31}|/E)\frac{\cos 2\theta_{13}}
   {\sqrt{2} G_{F}(\frac{Y_{e}}{m_{n}})(\frac{c_{0}}{R_{0}^{3}})}]^{\frac{1}{n_{h}}} 
   R_{0} Z_{0}^{(1+\frac{3}{n_{h}})},
\end{eqnarray}
where $Z_{0}=0.03$, $R_{0}=6.96 \times 10^{10}$ cm, 
and $F_{h} \sim 1$ (for small $\theta_{13}$).
Thus, how the uncertainty in $n_{h}$ alters the predicted bound for $\theta_{13}$
is regulated by the function $g(n_{h},\theta_{13})$, and
depends on whether the parameters $n_{h}$ and $\theta_{13}$
result in $g(n_{h},\theta_{13}) <1$ or $g(n_{h},\theta_{13})>1$.  The similar properties for
$\bar{P}_{nor}$ and $\bar{P}_{inv}$ also can be derived from
Eqs. (7), (9), (11), and (13).
As a brief summary, one notes that with the formulation and the chosen input parameters,
(i) $P_{l}=0$, and $\bar{P}_{l} =0$, and (ii) $P_{h}=\bar{P}_{h}=1$ if $g(n_{h},\theta_{13}) <1$,
while $P_{h}=\bar{P}_{h}=0$ if $g(n_{h},\theta_{13}) >1$.

%%%%%%%%%%%%%%%%%%%%%%%%%%%%%%%%%%%%%%%

%%%%%%%%%%%%%%%%% fig2%%%%%%%%%%%%%%%%%%%
\begin{figure}
\caption{The adiabaticity parameters is modified for a finite resonance layer, 
within which the density
and the mass scale $c$ are both variable. The contours
for $g(n_{h},\theta_{13}) \equiv \frac{\pi}{2 \ln2}\gamma_{h}=1$ 
are shown in the $n_{h}-\theta_{13}$ space 
with $Z=Z_{0}$, $Z=Z_{0}(1+10\%)$, and $Z=Z_{0}(1-10\%)$, where $Z_{0}=0.03$.  
Note that since the estimated 
deviation ($<2.7\%$) is barely observable in the parameter space, an exaggerated
uncertainty of $10\%$ is shown here for the purpose of illustration and comparison.
We may conclude that
this small deviation does not alter the general picture of the analysis.  \label{f2}} 
\centerline{\epsfig{file=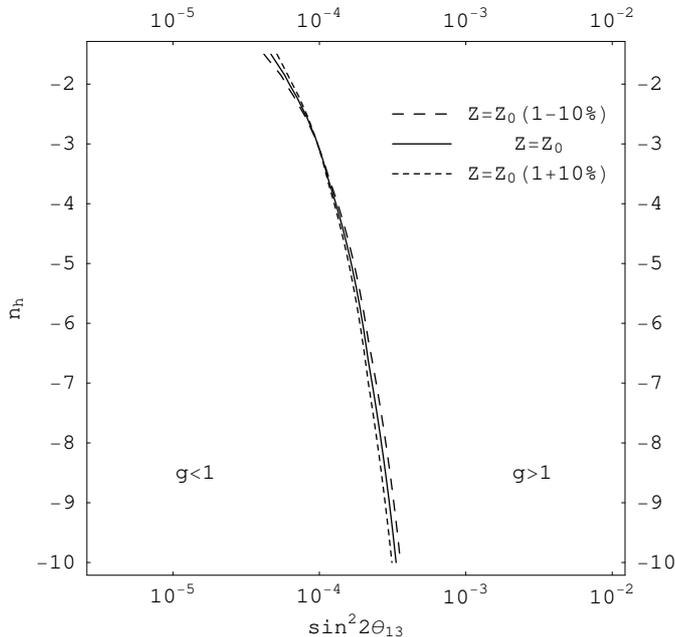,width= 9 cm}}
\vspace*{8pt}
\end{figure} 
%%%%%%%%%%%%%%%%%%%%%%%%%%%%%%%%%

In terms of the survival probabilities $P$ and $\bar{P}$ for 
$\nu_{e}$ and $\bar{\nu}_{e}$, respectively,
the uncertain local density profile, the possible mass hierarchies,
and the undetermined mixing angle $\theta_{13}$ 
together lead to three possible combinations of
($P$,$\bar{P}$), as indicated by $(a)$, $(b)$, and $(c)$ in Table~\ref{ta1}.  
In general, separating the scenarios is unlikely if 
$\sin^{2}2\theta_{13} <10^{-5}$ since $P_{nor}$ and 
$P_{inv}$, as well as $\bar{P}_{nor}$ and $\bar{P}_{inv}$, cannot be distinguished.  
In addition, there is a non-trivial dependence of $P$ and $\bar{P}$ on both $n_{h}$
and $\theta_{13}$ near $\sin^{2}2\theta_{13} \sim 10^{-4}$. 
This implies that establishing a better lower bound for $\theta_{13}$
relies on whether the uncertainty of $n_{h}$ in this region of parameter space can be reduced.
Based on Table~\ref{ta1}, we discuss the following cases for the purpose of illustration.
(i) If the mass hierarchy is identified as normal by the observation of extremely small $P$,
the predicted lower bound for $\theta_{13}$
could still be uncertain by more than one order of magnitude because of the uncertain $n_{h}$.
In this case, the condition
$g(n_{h},\theta_{13})>1$ establishes the predicted lower bound for $\theta_{13}$
with a specific value of $n_{h}$.  
(ii) On the other hand, a small $\bar{P}$ implies inverted hierarchy,
and the predicted bound for $\theta_{13}$ is also regulated by the condition  
$g(n_{h},\theta_{13})>1$. (iii) If both $P$ and $\bar{P}$ are observed to be large:
$P \sim 0.3$ and $\bar{P} \sim 0.6$, then it implies $g(n_{h},\theta_{13})<1$
with no information about the mass hierarchy.

%%%%%%%%%%%%%%%%%%%%%%%%%%%%%%%%%%%%%%%%%

Note that with the estimated location of resonance $Z_{0}=0.03$, 
there is a potential source of uncertainty in calculating the adiabaticity parameters.
The key point is, with the variations of $n_{h}$, $c$, and the density
in a finite resonance layer,
what modification to $\gamma_{h}$ is required?
We first show in Fig. 2 the contour (solid line) of $g(n_{h},\theta_{13})=1$
in the $n_{h}-\theta_{13}$ space for $Z_{0}=0.03$.
In order to estimate the uncertainty, one needs to calculate the width of the resonance layer 
\begin{equation}
\delta_{0} \simeq \frac{2 \tan 2\theta_{13}}{|\frac{1}{N_{e}}\frac{dN_{e}}{dr}|_{0}},
\end{equation}
which can be reduced to $\delta_{0} \simeq 2(0.03R_{0})\tan 2\theta_{13}/|n_{h}|$.
One notes from the solid contour of Fig. 2 that, as $n_{h}$ varies in the region
$-10<n_{h}<-1.5$, the higher resonance occurs roughly in the region
$5 \times 10^{-4} <\sin^{2} 2\theta_{13} <4 \times 10^{-3}$. 
This leads to an estimated bound for the width, $\delta_{0}/(Z_{0}R_{0}) <2.7\%$,
for the given $n_{h}$ and $\theta_{13}$ in the parameter space.
The deviation of the contour $g(n_{h},\theta_{13})=1$ caused by this small width
is barely observable in the $n_{h}-\theta_{13}$ space of Fig. 2.
To illustrate the smallness of this deviation and as a comparison, 
we also show in Fig. 2 the contours 
of $g(n_{h},\theta_{13})=1$ due to an exaggerated width of the resonance layer, regardless
of the values for any given $n_{h}$ and $\theta_{13}$.  
It is seen that even with the conservative estimation based on
the enlarged width $Z=Z_{0}(1\pm 10\%)$, 
the resultant deviation of the contour from that of $Z_{0}=0.03$
is still very small.  We may conclude that the small uncertainty due to
the finite width of the layer does not alter the general picture of the analysis.
Note that all the three contours meet at $n_{h}=-3$.
This can be realized from Eq. (16), in which the contribution
from any deviation of $Z_{0}$ vanishes when $n_{h}=-3$. 

%%%%%%%%%%%%%%%%%%%%%%%%%%%%%%%%%%%%%%%%%%%%%%%%%%%%%%%%%%%%%

The survival probabilities will be modified by the regeneration
effect as the neutrinos propagate through the Earth matter.
The three possible combinations $(a)$, $(b)$, and $(c)$ for ($P,\bar{P}$) are now modified
according to

\begin{eqnarray}\label{eq:effp}
P^{(a)}&=&\sin^{2}\theta_{13} P_{2e},  \nonumber \\
                   P^{(b)}&=&P_{2e},  \nonumber \\
P^{(c)}&=& P_{2e},
\end{eqnarray}
and
\begin{eqnarray}\label{eq:aeffp}
\bar{P}^{(a)}&=& 1-\bar{P}_{2e},  \nonumber \\
 \bar{P}^{(b)}&=& \sin^{2}\theta_{13} (1-\bar{P}_{2e}),  \nonumber \\
 \bar{P}^{(c)}&=& 1-\bar{P}_{2e}, 
\end{eqnarray} 
where $P_{2e}$ ($\bar{P}_{2e}$) is the probability that a 
$\nu_{2}$ ($\bar{\nu}_{2}$) arriving at the Earth surface is eventually detected
as a $\nu_{e}$ ($\bar{\nu}_{e}$) at the detector~\cite{p2e-1,p2e-2}.   %%%%%%%%%%

%%%%%%%%%%%%%%%%%%%%%%%%%%%%%%%%%%%%%%%%%%%%%%

%%%%%%%%%%%% Table 2 %%%%%%%%%%%%%%%%%%%

 \begin{table}
\caption{The average energies and relative luminosity for neutrinos of 
different flavors suggested by the G1, G2, and the LL models.}
 {\begin{tabular}{cccccc} \hline \hline  
     &  $\langle E^{0}_{\nu_{e}}\rangle$/MeV  & 
     $\langle E^{0}_{\bar{\nu}_{e}}\rangle$/MeV &     
  $\langle E^{0}_{\nu_{x}}\rangle$/MeV  & $L^{0}_{\nu_{e}}/L^{0}_{\nu_{x}}$  & 
   $L^{0}_{\bar{\nu}_{e}}/L^{0}_{\nu_{x}}$            \\ \hline
 LL & 12 & 15 & 24 & 2 & 1.6 \\
 G1 & 12 & 15 & 18 & 0.8 & 0.8 \\
 G2 & 12 & 15 & 15 & 0.5 & 0.5 \\
                        \hline
     \end{tabular}\label{ta2}}
 \end{table}
%%%%%%%%%%%%%%%%%%%%%%%%%%%%%%%%%%%%%%%%%%%%%%%%%%

%%%%%%%%%%%%%%%%% fig3%%%%%%%%%%%%%%%%%%%
\begin{figure}
\caption{The neutrino incident angles
at detectors $A$ and $B$, as measured individually from the local zenith,  
are given by $\psi_{\alpha}$ and $\psi_{\beta}$, respectively.
The angle formed by $A$, $O$ (the Earth center), and $B$ is denoted as
$\lambda$. \label{f3}} 
\centerline{\epsfig{file=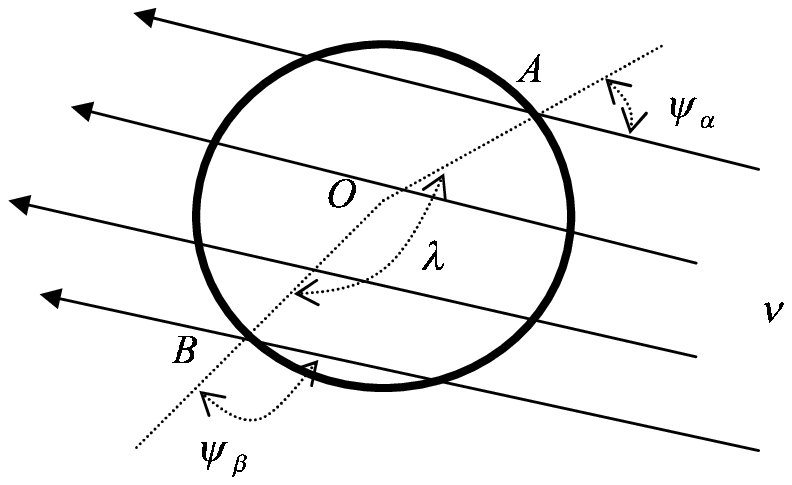,width=6 cm}}
\vspace*{8pt}
\end{figure} 
%%%%%%%%%%%%%%%%%%%%%%%%%%%%%%%%%

%%%%%%%%%%%%%%%%%%%  sec 4 %%%%%%%%%%%%%%%%%%%%%%%%%%%%%%%%%%%%%%%%%%%%%%%%%%%%%%

\section{SN Neutrino Spectral Parameters and Multi-detector Analysis}

In addition to the brief pulse of the neutronization $\nu_{e}$ burst,
all three flavors of neutrinos and antineutrinos are emitted from
the SN through pair production processes during a typical
time scale $\sim 10$s. 
The primary SN neutrino spectrum, which is not pure thermal, 
is usually modeled as a pinched Fermi-Dirac distribution
with several spectral parameters:

\begin{equation}\label{eq:f0}      
 F^{0}_{i} \sim \frac{L_{i}}{T^{4}_{i} f_{3}(\eta_{i})}
 \frac{E^{2}}{e^{[(E/T_{i})-\eta_{i}]}+1},
\end{equation} 
where $L_{i}$ is the luminosity of the neutrino flavor $\nu_{i}$, 
$T_{i}$ is the effective 
temperature of $\nu_{i}$ inside the respective neutrinosphere, $E$ is
the energy, $f_{3}(\eta_{i})$ is the normalization factor,  
and $\eta_{i}$ is the pinching parameter for $F^{0}_{i}$.
Notice that
$\eta_{\nu_{e}} \sim \eta_{\bar{\nu}_{e}} \sim \eta_{\nu_{x}} \sim 3.0$ 
is assumed here, where 
$\nu_{x}=\nu_{\mu}$, $\nu_{\tau}$, $\bar{\nu}_{\mu}$, and $\bar{\nu}_{\tau}$.
The most significant difference between the Garching models (G1, G2) and
the Lawrence Livermore model (LL) is that the Garching models
exhibit a less hierarchical structure in the luminosity and in the average energy
among different neutrino flavors, as shown in Table~\ref{ta2}.

We assume a core collapse SN at a distance 10 kpc away.
In a typical water Cherenkov detector, the isotropical inverse $\beta$-decay,
$\bar{\nu}_{e} + p \rightarrow n+e^{+}$, is the most dominant process.
In this work, other isotropical CC processes, 
$\nu_{e} + O \rightarrow F + e^{-}$ and $\bar{\nu}_{e} + O \rightarrow N + e^{+}$,
are also included in the calculation.  As for the directional events,
we include the scattering events induced by
$\nu_{i}$ and $\bar{\nu}_{e}$: $\nu_{i} (\bar{\nu}_{e})+ e^{-}$, where
$i=e, \mu, \tau$.
The cross sections for $\bar{\nu}_{e}+p$ and $\nu_{i}(\bar{\nu}_{e})$
are adopted from Ref.\cite{cross-1}, and that for $\nu_{e}(\bar{\nu}_{e})+O$
are adopted from Ref.\cite{cross-2}.
As shown in Fig. 3, 
the neutrino incident angle at a detector,
denoted as $\psi_{\alpha}$ and $\psi_{\beta}$ for detector
$A$ and $B$, respectively,
may be defined
as the angle measured from the local zenith.    
For a given $\psi_{\alpha}$, 
the allowed range for $\psi_{\beta}$ is limited.   %%%%%%%%%%%%%%%%%%%%
If the angle formed by
$A$, the Earth center, and $B$ is given by $\lambda$, then 
\begin{equation}
|\psi_{\alpha} - \lambda | < \psi_{\beta} < \psi_{\alpha} + \lambda
\end{equation}
for $\psi_{\alpha}+\lambda <\pi$, and 
\begin{equation}
| \psi_{\alpha}-\lambda | <\psi_{\beta} < 2\pi- (\psi_{\alpha}+\lambda)
\end{equation}
for $\pi < \psi_{\alpha}+\lambda < 2\pi$.
The density of Earth matter is of the order $\rho_{E} \sim 10 g/cm^{3}$,
and its variation is much smaller than that of the SN matter.
One may adopt a simple step function\cite{freund} as the approximation:   %%%%%%%%%%%%%%%%%%%%%%%
$\rho_{E} \approx 5.0$ $g/cm^{3}$ for $\frac{1}{2} r_{\oplus}<r<r_{\oplus}$ (mantle), 
and $\rho_{E} \approx 12.0$ $g/cm^{3}$ for $r< \frac{1}{2} r_{\oplus}$ (core), 
where $r_{\oplus}$ is the Earth radius. 
For the purpose of illustration, one may analyze the expected results
at two future detectors:
the HyperKamiokande\cite{hk} (detector $A$)  %%%%%%%%%%%%%%
and the MEMPHYS\cite{memphys} (detector $B$).  %%%%%%%%%%%%%%%%%
The locations of the two detectors lead to $\lambda \sim \pi/2$.

Since the neutrino flux 
penetrates significant amount of Earth matter 
only when the incident angle at a detector
is greater than $\pi/2$, no observable 
difference is expected from the two detectors
if both $\psi_{\alpha}$ and $\psi_{\beta}$
are less than $\pi/2$.  In this case, 
both $P_{2e}$ and $\bar{P}_{2e}$ reduce to $\sin^{2}\theta_{12}$,
and all the observed differences between $A$ and $B$ vanish.  
Given the variety of spectral models,
the present analysis is aiming at 
seeking for model-independent properties that would help identify
the working scenario.

%%%%%%%%%%%%%%%%%   sec 5  %%%%%%%%%%%%%%%%%%%%%%%%%%%%%%%%%%%%%%%%%%%

\section{Observables and Analyses}

%%%%%%%%%%%%%%%%fig 4%%%%%%%%%%%%%%%%%%%%%%%%%%

\begin{figure}
\caption{The ratio $\bar{D}^{(a)}$ as a function of the neutrino energy, with
$\psi_{\alpha}=\pi/4$ (detector $A$ is unshadowed) 
and $\psi_{\beta}=0.95 \pi$ (detector $B$ is shadowed by mantle and core).
The primary neutrino spectra are adopted from G1, G2, and LL models. \label{f4}} 
\centerline{\epsfig{file=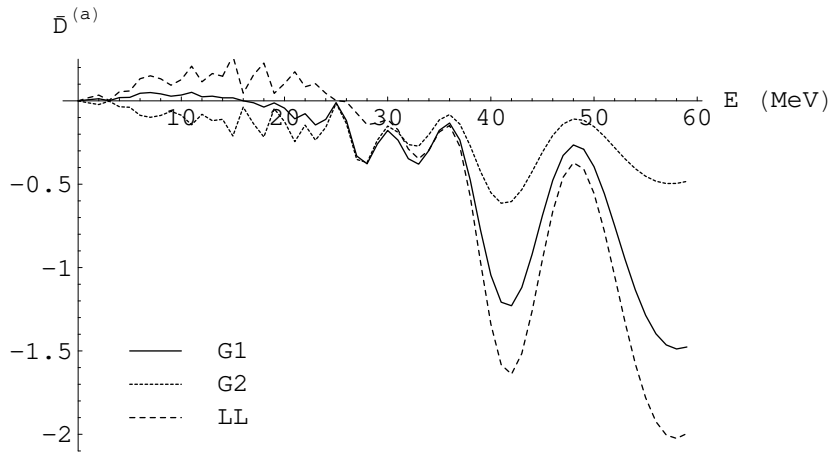,width=11 cm}}
\vspace*{8pt}
\end{figure}

%%%%%%%%%%%%%%%%%%%%%%%%%%%%%%%%%%%%%%%%%%%%%%%%%%%%%%%%%%%%%%

Simultaneous observation of the SN neutrinos 
at two terrestrial detectors would yield the most useful 
results when pronounced Earth regeneration effect is observed at only 
one of the detectors.   
Given the two terrestrial detectors $A$ and $B$, 
a proper analysis of the expected neutrino event rates at
the two detectors would in principle reveal the signatures representing
distinct scenarios of the neutrino properties and mixing schemes.  
To pave the way for the analysis, we first examine the energy dependence
of the relative flux difference at the detectors $A$ and $B$:

\begin{equation}
D = \frac{F^{B}_{\nu_{e}}-F^{A}_{\nu_{e}}}{F^{A}_{\nu_{e}}}
\end{equation}
for the $\nu_{e}$ flux, and
\begin{equation}
\bar{D} = \frac{F^{B}_{\bar{\nu}_{e}}-F^{A}_{\bar{\nu}_{e}}}{F^{A}_{\bar{\nu}_{e}}}
\end{equation}
for the $\bar{\nu}_{e}$ flux.  With $F^{0}_{\nu_{i}}$ and $F^{0}_{\bar{\nu}_{i}}$ denoting
the primary flux of $\nu_{i}$ and $\bar{\nu}_{i}$, respectively,
one may relate the observed and the primary fluxes as

\begin{equation}\label{eq:f}
F_{\nu_{e}}=F^{0}_{\nu_{e}}+(1-P)(F^{0}_{\nu_{x}}-F^{0}_{\nu_{e}}),
\end{equation}
\begin{equation}\label{eq:fa}
F_{\bar{\nu}_{e}}=F^{0}_{\bar{\nu}_{e}}+(1-\bar{P})(F^{0}_{\bar{\nu}_{x}}-F^{0}_{\bar{\nu}_{e}}).
\end{equation} 
The three scenarios then lead to several distinct relative flux differences: 
\begin{equation}
D^{(a)}=\frac{\sin^{2}\theta_{13}[F^{0}_{\nu_{e}}-F^{0}_{\nu_{x}}]
[P_{2e}-\sin^{2}\theta_{12}]}
{F^{0}_{\nu_{e}}[\sin^{2}\theta_{13}\sin^{2}\theta_{12}]+
F^{0}_{\nu_{x}}[1-\sin^{2}\theta_{13}\sin^{2}\theta_{12}]},
\end{equation}

\begin{equation}
D^{(b)}=\frac{[F^{0}_{\nu_{e}}-F^{0}_{\nu_{x}}]
[P_{2e}-\sin^{2}\theta_{12}]}
{F^{0}_{\nu_{e}}\sin^{2}\theta_{12}+
F^{0}_{\nu_{x}}\cos^{2}\theta_{12}},
\end{equation}

\begin{equation}
D^{(c)}=\frac{[F^{0}_{\nu_{e}}-F^{0}_{\nu_{x}}]
[P_{2e}-\sin^{2}\theta_{12}]}
{F^{0}_{\nu_{e}}\sin^{2}\theta_{12}+
F^{0}_{\nu_{x}}\cos^{2}\theta_{12}},
\end{equation}

\begin{equation}
\bar{D}^{(a)}=\frac{-[F^{0}_{\bar{\nu}_{e}}-F^{0}_{\bar{\nu}_{x}}]
[\bar{P}_{2e}-\sin^{2}\theta_{12}]}
{F^{0}_{\bar{\nu}_{e}}\cos^{2}\theta_{12}+
F^{0}_{\bar{\nu}_{x}}\sin^{2}\theta_{12}},
\end{equation}

\begin{equation}
\bar{D}^{(b)}=\frac{-\sin^{2}\theta_{13}[F^{0}_{\bar{\nu}_{e}}
-F^{0}_{\bar{\nu}_{x}}][\bar{P}_{2e}-\sin^{2}\theta_{12}]}
{F^{0}_{\bar{\nu}_{e}}[\sin^{2}\theta_{13}\cos^{2}\theta_{12}]+F^{0}_{\bar{\nu}_{x}}
[1-\sin^{2}\theta_{13}\sin^{2}\theta_{12}]},
\end{equation}

\begin{equation}
\bar{D}^{(c)}=\frac{-[F^{0}_{\bar{\nu}_{e}}-F^{0}_{\bar{\nu}_{x}}]
[\bar{P}_{2e}-\sin^{2}\theta_{12}]}
{F^{0}_{\bar{\nu}_{e}}\cos^{2}\theta_{12}+
F^{0}_{\bar{\nu}_{x}}\sin^{2}\theta_{12}}.
\end{equation}

It is seen that the ratios also depend on the chosen model
for the primary $\nu_{i}$ and $\bar{\nu}_{i}$ fluxes, as well as on $P_{2e}$ and
$\bar{P}_{2e}$, which introduce the non-trivial energy dependence
into $D$ and $\bar{D}$.  As an illustration, we show the variation of
$\bar{D}^{(a)}$ as a function of energy in Fig. 4, with
$\psi_{\alpha}=\pi/4$ (detector $A$ is unshadowed) 
and $\psi_{\beta}=0.95 \pi$ (detector $B$ is shadowed by mantle and core). 
The primary neutrino spectra are adopted from the G1, G2, and the LL models.
Note that with the finite resolution power of the detector,
not all the details of the curve can be observed in practice.
It is clear from Eq.(30) that the energy dependence of
$\bar{D}^{(a)}$ originates from the factor $\bar{P}_{2e}-\sin^{2}\theta_{12}$.
The energy dependence of $D$ and $\bar{D}$, as shown in 
Eqs.(27-32), can be used to illustrate the properties
of the observables that shall be proposed in this work.

%%%%%%%%%%%%%%%%% fig5%%%%%%%%%%%%%%%%%%%
\begin{figure}
\caption{The possible range of the value for $R=R(E)$ is indicated by
the vertical range between two horizontal lines. Results for scenarios (a), (b), and (c)
are shown here
with the chosen bin size of 5 MeV.  
The incident angles $\psi_{\alpha} \leq \pi/2$ 
(detector $A$ unshadowed) and 
$\psi_{\beta}=0.95 \pi$ (detector $B$ shadowed by mantle and core) are used
to optimize the result.  The estimated ranges of $R$ shown in the figures
are determined by the results of the three spectral models and the 
statistical errors. \label{f5}} 
\centerline{\epsfig{file=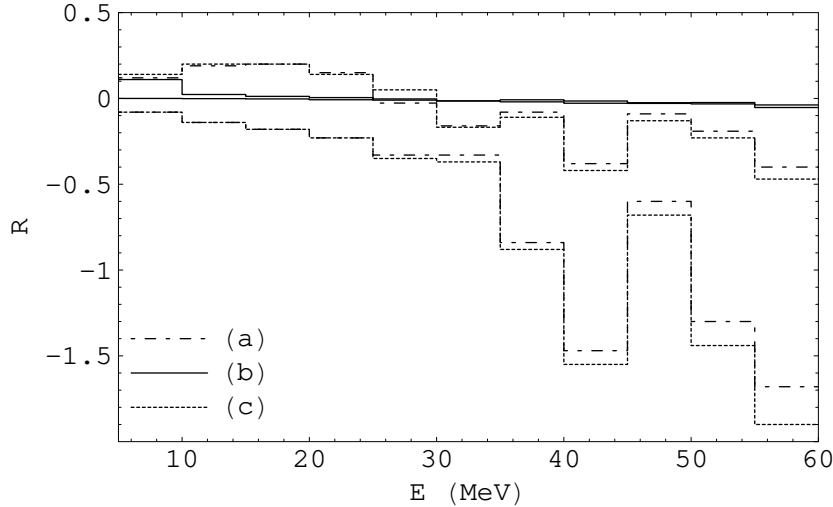,width=11.0cm}}
\vspace*{8pt}
\end{figure} 
%%%%%%%%%%%%%%%%%%%%%%%%%%%%%%%%%

\subsection{Relative difference of the event rates}

In terms of the event rates, one may first examine
the relative  
difference of the observed event rates per unit target mass for detectors $A$ and $B$:
 \begin{equation}
R \equiv \frac{N_{B}-N_{A}}{N_{A}}.
\end{equation}
The total event rates per unit target mass, $N_{A}$ and $N_{B}$, 
consist of all the isotropical and directional events mentioned
in the previous section, with
\begin{equation}
N_{A,B}=\frac{\sum N_{i}}{M_{A,B}},
\end{equation}
where $M_{A,B}$ denotes the total target mass of a detector.
The expected 
event rates $N_{i}$ induced by the flavor $\nu_{i}$ (or $\bar{\nu}_{i}$) at
the detector is given by
\begin{equation}
N_{i}=\frac{N_{t}}{4\pi L^{2}} \int [\sum_{y} F_{\nu_{i}} \times \sigma_{iy}]dE,
\end{equation}
where $N_{t}$ is the target number at the detector, $L$ is the distance to the supernova,
$\sigma_{iy}$ is the cross section for $\nu_{i}$ (or $\bar{\nu}_{i}$) in 
a particular reaction channel $y$.  The detection efficiency is assumed to be one. 
In the following, 
one may denote the resultant values of $R$ 
for the six scenarios as $R^{J}$,
where $J=(a), (b)$, $(c)$.

%%%%%%%%%%%%%%%%%%%%%%%%%%%%%%%%%%%%%%%%%%%%%%%%%%%%%%%%%%%

As an illustration, we show in Fig. 5 the expected values of $R$
for scenarios $(a)$, $(b)$, and $(c)$.
The reasonable estimation of the energy bin size 
is roughly a few MeV for the detection of typical SN neutrinos\cite{fogli:jcap05}. 
We adopt an energy bin size of 5 MeV in the figure.
Note that for scenario $(b)$, the deviation of $R$ from zero is
$\sim 1\%$ or smaller, and the sign of $R$ is undetermined with the error bar.
The estimated width for $R$ is given by the different  
results from the three spectral models and the statistical errors from each.
As an example, one notes that the expected values of $R^{(a)}$ under LL, G1, and G2 models
in 25 MeV $< E <$30 MeV are estimated to be $-0.085 \pm 0.057$, $-0.221 \pm 0.112$,
and $-0.216 \pm 0.109$, respectively. The resultant $R^{(a)}$
is then shown in Fig. 5 with the combined width $-0.333 < R^{(a)} < -0.027$
for 25 MeV $< E <$30 MeV.
Note that there is no apparent energy dependence of the width.

Even though some details of the energy-dependence 
in Fig. 5 are lost, certain
useful qualitative properties are still available.
It is seen that the Earth effect could suppress
the event rates to an observable 
level (of order 10\% or more) for $E \geq 25$ MeV if the working scenario is $(a)$ or $(c)$.
The sign and the magnitude of $R$ in $E \geq 25$ MeV can be used to separate the three
scenarios into two groups: $[(a),(c)]$ and $(b)$.
It should be pointed out that for $E<25$ MeV, 
the uncertainties wipe out the small deviation from zero
and the sign of $R$ is undetermined for all scenarios.   

%%%%%%%%%%%%%%%%% fig6%%%%%%%%%%%%%%%%%%%
\begin{figure}
\caption{The values of $S=S(E)$ estimated  
with the bin size of 5 MeV and $\psi_{\alpha} \leq \pi/2$ (detector $A$ unshadowed), 
$\psi_{\beta}=0.95 \pi$ (detector $B$ shadowed by mantle and core). 
The ranges of $S$ shown in the figure
are based on the 
combined results of the three spectral 
models and the statistical errors. \label{f6}} 
\centerline{\epsfig{file=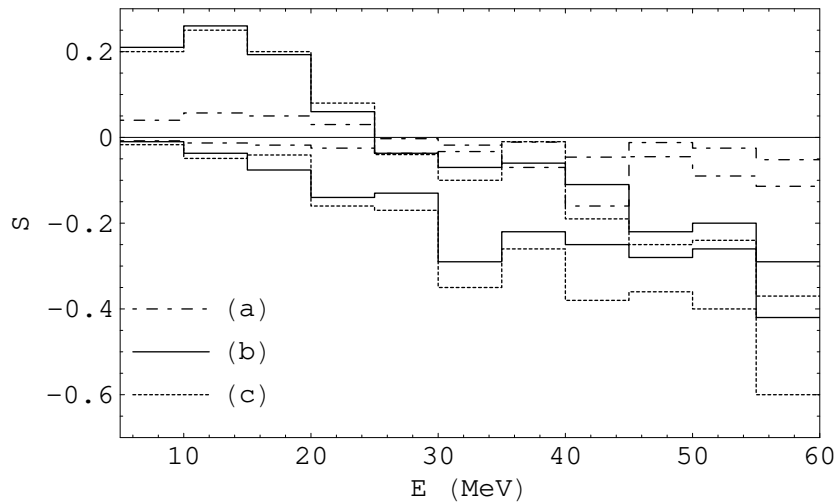,width=11.0 cm}}
\vspace*{8pt}
\end{figure} 
%%%%%%%%%%%%%%%%%%%%%%%%%%%%%%%%%

More hints may be available from analyzing the
directional events alone,
which are predominantly the $\nu_{e}$-induced
elastic scattering events.  However, 
the directional event rate is approximately $1 \sim 2$ orders of magnitude
less than the isotropical event rate.  
Extracting the directional events effectively from the dominant background
of the isotropical events may be quite challenging in practice.
In the following analysis,
one assumes that the directional event rates 
in the forward cone with a narrow solid angle can be identified.

\subsection{Relative difference of the directional events}

We may define another ratio based on the directional events
per unit mass of target at the two detectors,
\begin{equation}
S \equiv \frac{n_{B}-n_{A}}{n_{A}},
\end{equation}
where $n_{B}$ and $n_{A}$ denote the shadowed and the unshadowed
directional events per unit target mass, respectively.  
The properties of $S$, as shown in Fig. 6, lead to new information that is unseen
from $R$. 
One notes that two groups of scenarios, $(a)$ and $[(b),(c)]$,  
could be identified by the magnitude of $S$ at the high energy end
since the magnitude could differ by a factor of three or more.  
Thus, one may conclude that
the degeneracy between $R^{(a)}$ and $R^{(c)}$ could be
removed by using the observable $S$, and that the three scenarios
could be reasonably distinguished by analyzing both $R$ and $S$ at the high energy end.

%%%%%%%%%%%%%%%%%%%%%%%%%%%%%%%%%%%%%%%%%%%%%%%%%%%%%%%%%%%

It should be pointed out that even though $R$ and $S$ represent energy-integrated quantities,
they still exhibit general energy dependence with the chosen finite bin size,
as can be seen from Figs. 5 and 6.
To qualitatively illustrate this general properties,
one may approximate $R$, for simplicity, as the event ratios that consist  
only of the predominant $\bar{\nu}_{e} + p$ events. 
For scenario $(a)$, Eqs.(25-26) and (33-35) lead to
\begin{equation}
R^{(a)} \simeq \frac{[F^{0}_{\bar{\nu}_{e}}-F^{0}_{\bar{\nu}_{x}}] 
\int (\bar{P}_{2e}-\sin^{2} \theta_{12}) \sigma (E) dE}
{\int (F^{0}_{\bar{\nu}_{e}} \cos^{2} \theta_{12} + F^{0}_{\bar{\nu}_{x}} \sin^{2} \theta_{12})
\sigma (E) dE},
\end{equation}
where $\sigma (E)$ is the cross section.
With the input parameters and a given model for the fluxes,
\begin{equation}
R^{(a)} \sim \frac{\int (\bar{P}_{2e}-\sin^{2} \theta_{12}) \sigma (E) dE}
{\int \sigma (E) dE}. 
\end{equation}
If one calculates $R^{(a)}$ in a relative small step in energy, $\Delta E \sim$ a few MeV,
the cross section $\sigma = \sigma (E)$, which is roughly $\sigma \sim E^{2}$,
would be relatively smooth as compared to the function $\bar{P}_{2e}$.
As a consequence, one may expect $R^{(a)} \sim \bar{P}_{2e}-\sin^{2} \theta_{12}$.
It is seen from Eq.(30) that the qualitative behavior of the factor,
$\bar{P}_{2e}-\sin^{2} \theta_{12}$, which dictates the general trend of the energy dependence of
$\bar{D}^{(a)}$ in Fig. 4, propagates to $R^{(a)}$ in Fig. 5.
The energy dependence of $S$, and that of $T$, which is introduced in the next subsection,
can be realized in a similar way.

\subsection{Double ratio with both directional and isotropical events}

The observables,
$R$ and $S$, tend to flip signs near the peak of the spectrum ($\sim 25$ MeV),
and become less useful for analyzing the abundant events induced by the neutrinos
with energies near the peak.
It would be worth while
to further investigate whether there is more to learn from 
the middle range of the spectrum.
Theoretically, the Earth regeneration effect is signaled
by the oscillatory modulations of
the observed spectra.  A quantitative analysis of the 
resultant event rates, however, may be quite subtle.
The resolution power of the detector is limited and
the modulation may not alter the total event rates significantly enough to
render an observable difference between the shadowed and 
the unshadowed events. To optimize the results for an analysis, 
neutrino event rates from a specific range of 
energy cut may be needed.  

%%%%%%%%%%%%%%%%% fig7%%%%%%%%%%%%%%%%%%%
\begin{figure}
\caption{The possible ranges of $T=T(\psi_{\beta})$ for scenarios $(a)$, $(b)$, 
and $(c)$ are estimated
from the combined results of the three spectral models
and the statistical errors for $15 <E < 30$ MeV. The incident angle
for detector $A$ is chosen as $\psi_{\alpha}=\pi/2$, and an assumed
directional resolution of $\pi/12$ for detector $B$ is adopted.
The resultant range of $T$ in each scenario is 
expected to show no 
$\psi_{\beta}$-dependence for $\psi_{\beta} \le \pi/2$. \label{f7}} 
\centerline{\epsfig{file=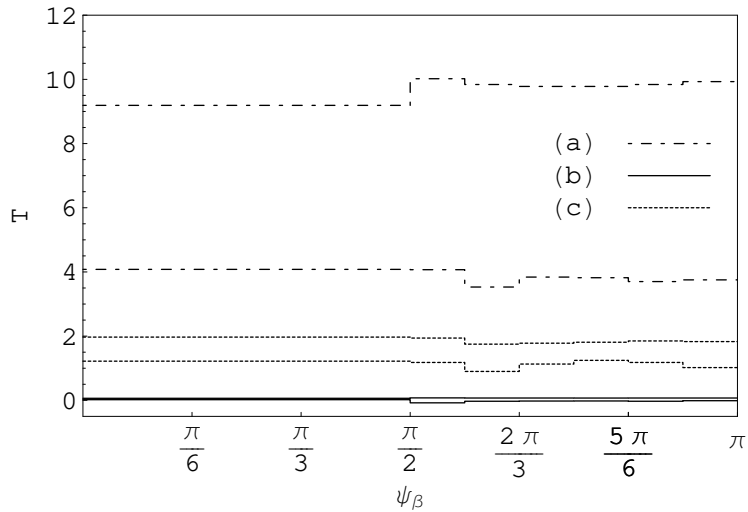,width=11 cm}}
\vspace*{8pt}
\end{figure} 
%%%%%%%%%%%%%%%%%%%%%%%%%%%%%%%%%
%%%%%%%%%%%%%%%%%%%%%%%%%%%%%%%%%%%%%%%%%%%%%%%%%%%%%%%%%%%%%%%%%%%%%%%

One may construct a double ratio with both the isotropical
and the directional events:
\begin{equation}
T \equiv \frac{[(N_{B}-n_{B})-(N_{A}-n_{A})]/(N_{A}-n_{A})}{(n_{B}-n_{A})/n_{A}},
\end{equation}
where $N-n$ and $n$ denote the isotropical and the directional events,
respectively.  Depending on the time of a day,
a given incident angle $\psi_{\alpha}$ for detector $A$ corresponds to a definite
incident angle $\psi_{\beta}$ for detector $B$.  
Notice that the possible values of $\psi_{\beta}$ are regulated by Eqs. (21) and (22).
If $\psi_{\alpha} \sim 0$
or $\psi_{\alpha} \sim \pi$, the angle $\psi_{\beta}$ 
is limited to a narrow range: $\psi_{\beta} \sim \lambda \sim \pi/2$. 
On the other hand
if $\psi_{\alpha} \sim \pi/2$, the allowed range for $\psi_{\beta}$
becomes maximized: $0< \psi_{\beta} <\pi$, and
the observables are expected to vary with a wide range of $\psi_{\beta}$.
Each of the two detectors
may be unshadowed (if $\psi_{\alpha, \beta} < \pi/2$), 
shadowed by the mantle (if $\pi/2 < \psi_{\alpha, \beta} < 5\pi/6$), 
or shadowed by the mantle and the core (if $5 \pi/6 < \psi_{\alpha, \beta} < \pi$).
In any case, more useful hints from a simultaneous observation would be available if 
one of the incident angles is greater than $\pi/2$ and the other 
is less that $\pi/2$.

The observable $T$ may be used as a supplementary tool
to the other two observables, $R$ and $S$. 
As an illustration, the ratio $T$ is evaluated here for $15 <E < 30$ MeV
and analyzed as a function of $\psi_{\beta}$, with
$\psi_{\alpha}=\pi/2$ and a conservatively assumed
directional resolution of $\pi/12$ for detector $B$, as shown in Fig. 7.
Notice that the pointing accuracy of the incident neutrinos
at the SK detector can reach the level to 
within a few degrees at 10 kpc~\cite{tomas:03}.   
The results for each scenario are expected to show no 
$\psi_{\beta}$-dependence for $\psi_{\beta} \le \pi/2$.  
In addition, calculations show that scenarios $(b)$ leads to 
small deviations from zero: $-0.078<T<0.076$, which can be 
distinguished from the other two scenarios.  
Furthermore, $T^{(a)}$ and $T^{(c)}$ can also be distinguished by the 
difference in magnitude.
One would expect that a combined analysis of $R$, $S$, and $T$
should be able to provide certain useful constraint or prediction about
the neutrino parameters.

%%%%%%%%%%%%%%%%%% sec 6 %%%%%%%%%%%%%%%%%%%%%%%%%%%%%%%%%%%

\section{Summary and Conclusions}
 
The SN neutrino burst has long been considered as a promising
tool for probing the 
neutrino mass hierarchy and the mixing angle $\theta_{13}$.
The current consensus is that 
the contributions from both
the MSW and the non-MSW effects should be included
in analyzing the flavor transition of the SN neutrino signals.
However, the uncertain local density profile 
of the SN matter and the neutrino spectral parameters
may lead to diverse interpretations to the neutrino signals
from the contribution of the MSW effect.
To investigate the consequences resulting 
from these uncertainties, this work 
suggests a more general parametrization for the uncertain local density profile, and
outlines a scheme using ratios of experimental observables as the
discriminators for the possible scenarios that are related to the MSW effect.
As an illustration, we analyze the expected  
event rates at two terrestrial Cherenkov detectors. 
The consequences due to choices of the existing spectral models 
are included in the analysis. 
The possible scenarios arising from the uncertainties are then examined
through the observables $R$, $S$, and $T$. 
It is seen that the uncertain local density profile of the SN matter can alter
the predictions for $\theta_{13}$ and the mass hierarchy from the contribution
of the MSW effect alone.  The predicted bound for $\theta_{13}$ is regulated by the
proposed function $g(n_{h},\theta_{13})$.

The results in this work are derived from the expected event rates at two
megaton-scale Cherenkov detectors that would be available in the future.
Analyses of the future SN neutrinos based on other types of detectors,
such as the liquid Argon and kton-scale scintillation detectors
that are sensitive to distinct channels of the neutrino interactions,
should also provide valuable information toward a better understanding
of the neutrino parameters and the SN physics.  A detailed analysis based on all the three types
of detector is performed in, $e.g.$, Ref. 6, in which seven SN and neutrino 
parameters are taken into consideration.  To extend the present work, 
it would be intriguing to investigate how the uncertain density profile would impact
the outcome of the analysis based on the detectors other than the Cherenkov ones.

In shaping a possible new paradigm for the neutrino flavor 
conversion in a SN, recent studies and simulations
suggest that the shock wave propagation and certain types of non-MSW flavor conversion
may occur in the complex environment of a 
SN at different space and time scales if the proper 
physical conditions are met.
These effects, such as the neutrino 
self coupling in dense 
media\cite{raffelt:07,dd:08,flmm:08,fq:06,hr:06,df:06,pr:02,rs:07,ad:08,ddm:08} %%%%%%%%%%%%%%%% 
and the neutrino flavor de-polarization associated with
the after shock turbulence\cite{benatti:95,fg:06,mon:06,chou:07}, %%%%%%%%%%
may have different origins from that of the MSW effect and
would impact the efficiency of neutrino flavor transition.

It is suggested that the neutrino self interactions induce
collective flavor transitions, which may occur before
or in the same region as the resonant flavor transition.
The size of modification to the original neutrino flux depends on
the mass hierarchy, as well as on the mixing angle $\theta_{13}$.
As an example, one considers the simplest case when the self-interaction 
effects can be factored out
from the ordinary MSW effects.  For the inverted hierarchy, 
it is suggested that if $\theta_{13}$ is nonzero,
the collective pair-conversion
of the type $\nu_{e}\bar{\nu}_{e} \rightarrow \nu_{x}\bar{\nu}_{x}$
can be triggered 
before the ordinary MSW effect occurs\cite{lisi:07}.  %%%%%%%%%%%%%%%%%%%%%%%%%%%%%%%%%%%%%%
Thus, even in the simplest case, the altered primary spectra
could impact the validity and the usefulness
of the observables proposed in this work. 
A better understanding of the modified neutrino spectra resulting from the collective effects
would help in establishing more convenient and useful observables
since the knowledge of the neutrino spectra before and after the 
MSW effect is essential for the present analysis. 
Note that the treatment of the variable density profile for the MSW effect,
as that proposed in this work,
is unaffected by the non-MSW effects.

The shock wave
propagation, on the other hand, may alter
the density profile of the SN matter and
induce certain non-linear effects that modify the neutrino survival
probabilities and the neutrino spectrum. 
In particular, there is a possibility that more than one MSW-resonance
related to the same scale of mass-squared difference could arise due
to the shock wave propagation.
The effects of the shock wave on certain physical observables,
with the emphasis on the time-dependent properties,
are analyzed\cite{Ga,fogli:jcap05,sf}.
Although the proposed scheme in the present work is limited to
the discussion without considering specific scenario due to the shock effects,
it takes into account the possible varying density profile,
and is not limited to whether the profile is actually primary.
In addition, the quantities $R$, $S$, and $T$
are proposed in this work to examine the time-integrated behaviors of the event rates.
As an extension to this present analysis,
it would be intriguing to further investigate weather
the time structure of the shock waves effects and
the possible multiple MSW-resonance due to the shock effect can
alter the time-integrated properties of the proposed quantities to an observable level.
Even with the unknown neutrino parameters, 
the uncertain model for the SN neutrino spectrum, and the possible statistical
errors, the quantities $R$, $S$, and $T$ might still be able to provide
another potential means for probing the SN physics if they reveal time-independent
signatures that are 
related to the shock wave propagation.
On the other hand,
if the shock signatures are unobservable with these time-integrated quantities, 
then these quantities
might be able to provide insights to the neutrino parameters without being
affected by the shock waves.

%%%%%%%%%%%%%

It should be emphasized that
the present work is devoted to effects that are
related to the MSW resonant flavor transition in a SN.
The focus is aimed at the possible modification to the
results of MSW effect due to the uncertain density profile.  
This modification is expected to have certain impact to the complete analysis based on both
the MSW effect and the non-MSW effects.
The whole details of the non-MSW effects,
as well as the interplay between  
the MSW and the non-MSW consequences, are still not fully understood.
The neutrino flavor conversion in a SN remains as
a complicated problem that awaits for more complete answers.
Whether the flavor transition arising from one origin
would dominate over that from the others depends on 
the uncertain spectral parameters, the undetermined neutrino intrinsic properties,
and the physical conditions that the neutrinos encounter at 
different stages of the SN environment.
In addition, the reliability of the analysis relies on whether
the MSW and non-MSW effects can be analyzed separately.
However, these unrelated types of effects decouple only
when specific physical conditions are met.

In any case, analysis of the non-trivial MSW flavor conversion due to the
possible drastic variation of the local density profile,
as discussed in this work,
should provide certain insight in probing the neutrino parameters and in
shaping a more convincing paradigm for the neutrino
flavor conversion in the SN. 

\acknowledgments This work is supported by the National 
Science Council of Taiwan under grant No. NSC97-2112-M-182-001. 

%%%%%%%%%%%%%%%%%%%%%%%%%%%%%%%%%%%%%%%%%%%%%%%%%%%%%%%%%%%%%%%%%%%%%%

\end{document}